# A multicore parallel algorithm for multiscale modelling of an entire human blood circulation network

Jiawei Liu[a][†] and Hiroshi Suito[a]*

[a] *Advanced Institute for Materials Research, Tohoku University, Sendai, Japan*

[†] liu.jiawei.e2@tohoku.ac.jp

* hiroshi.suito@tohoku.ac.jp

## ABSTRACT

The presented multi-scale, closed-loop blood circulation model includes arterial, venous, and portal venous systems, heart-pulmonary circulation, and micro-circulation in capillaries. One-dimensional models simulate large blood vessel flow, whereas zero-dimensional models are used for simulating blood flow in vascular subsystems corresponding to peripheral arteries and organs. Transmission conditions at bifurcation and confluence are solved using Riemann invariants. Blood circulation simulation in the portal venous system and related organs (liver, stomach, spleen, pancreas, intestine) is particularly targeted. Those organs play important roles in metabolic system dynamics. The proposed efficient parallel algorithms for multicore environments solve these equations much faster than serial computations.

## 1. Introduction

The cardiovascular system, an intricate network including the heart, large arteries, arterioles, peripheral arteries, capillaries, venules, and large veins, has been attracting intense interest among researchers in medicine, biology, physics, and mathematics. Nevertheless, the mysteries of the system have not yet been understood adequately. The blood circulating through the network is transporting oxygen, carbon dioxide, and widely diverse nutrients. That circulation plays a fundamentally important role in homeostasis in the human body. To describe and simulate the dynamics in such a complicated network, reduced-order models such as one-dimensional (1D) and zero-dimensional (0D) models have been constructed (Sun et al. 1997; Formaggia et al. 2003; Quarteroni and Formaggia 2004; Milišić and Quarteroni 2004; Formaggia et al. 2009; Müller and Toro 2014; Quarteroni et al. 2019; Müller et al. 2023). Numerous research efforts, from less complicated models to more complicated specific or holistic models, have been presented and have been applied widely to elucidate the targeted phenomena, describing, for instance, pulses and waves in arteries (Anliker et al. 1971), ventricular–arterial coupling during aging (Liang et al. 2009), thermo-fluid activity of the upper limb (He et al. 2004), cerebrospinal fluid dynamics (Toro et al. 2022), and heart models (Formaggia et al. 2006; Quarteroni et al. 2017; Regazzoni et al. 2022). Providing appropriate upstream and downstream boundary conditions in three-dimensional (3D) blood flow simulations for detailed local geometries is another useful application of multiscale models (Chi et al. 2022).

Although arterial and venous systems are two main components of our circulatory system, another essential component is the hepatic portal vein, which has been included in models reported by Mynard and Smolich (2015) and by Audebert et al. (2017). The portal vein transports nutrients and toxins from the intestine and hormones from the pancreas to the liver. Another input to the liver, the hepatic artery, makes the portal venous system rather complicated. As reported herein, we constructed a closed-loop global 1D-0D model including the portal venous system to examine the blood flow there, which governs the transport processes of nutrients and hormones among abdominal organs. A fluid dynamical characteristic of the portal vein is that both ends are capillaries and are not connected directly to the heart. Therefore, the flow is not so strong. It is affected sensitively by the behaviors of the 0D models at both ends, representing capillaries, which causes very slow convergence in iterations. Therefore, we have constructed a parallel implementation using a minimization problem.

This paper is organized as follows. In Section 2, we introduce the mathematical models used to treat each compartment in the closed-loop circulation, including the arterial, venous, and portal venous systems, micro-vasculature, and heart–pulmonary circulation. In Section 3, numerical schemes used to compute the solution of the equation systems along with the transmission conditions for bifurcation/confluence between 1D segments and coupling condition are described. In Section 4 and Section 5, computational results and the detailed parallel computation process are presented, with subsequent discussions in Section 6. Section 7 presents conclusions and future prospects.

## 2. Mathematical models of the blood circulation network

To represent blood circulation in the human body comprehensively, it is necessary for our mathematical model to incorporate blood flow through the arterial system, venous system, and portal venous system, and to connect those flows appropriately by the associated micro-vasculature and heart–pulmonary circulation. The following introduces details of the mathematical models used to simulate the blood flow in these compartments.

### *2.1 Modeling of the arterial and venous networks*

In overall blood circulation, blood enters the aorta, which is the largest artery and which is connected to the heart. It then flows through the arterial network with numerous bifurcations, reaching all the regions of the body (Figure 1(a)). Then, via nodes representing the micro-vasculature, it enters the venous network with confluence and converges to the vena cava (Figure 1(b)). Here and in the following discussion, a bifurcation is defined as the location at which blood separates into two vessels. The term confluence is its opposite.

Table 1. Vessel names for arteries and veins. Numbers at the left correspond to segments shown in Figure 1.

| | | | |
|---|---|---|---|
| 1 Ascending aorta | 28 R. posterior cerebral artery I | 55 L. posterior tibial artery | 82 L. ulnar vein II |
| 2 Aortic arch I | 29 L. anterior cerebral artery II | 56 Splenic artery | 83 L. interosseous vein |
| 3 Brachiocephalic | 30 R. anterior cerebral artery II | 57 Gastric artery | 84 L. radial vein |
| 4 Thoracic aorta I | 31 Anterior communicating artery | 58 Hepatic artery | 85 L. ulnar vein I |
| 5 L. common carotid | 32 L. posterior cerebral artery II | 59 Mesenteric Portal Vein | 86 R. subclavian vein |
| 6 R. common carotid | 33 R. posterior cerebral artery II | 60 Splenic portal vein | 87 L. subclavian vein |
| 7 R. subclavian artery I | 34 Celiac artery | 61 Hepatic Portal Vein I | 88 Inferior sagittal sinus |
| 8 Thoracic aorta II | 35 Abdominal aorta I | 62 Gastric portal vein | 89 internal cerebral vein |
| 9 L. subclavian artery I | 36 Mesenteric artery | 63 Hepatic Portal Vein II | 90 R. internal jugular vein |
| 10 L. external carotid artery | 37 Abdominal aorta II | 64 Hepatic vein I | 91 Superior sagittal sinus & L. internal jugular vein |
| 11 L. internal carotid artery I | 38 R. common iliac artery | 65 Hepatic Vein II | 92 R. external jugular vein |
| 12 R. internal carotid artery I | 39 L. common iliac artery | 66 R. posterior tibial vein | 93 L. external jugular vein |
| 13 R. external carotid artery | 40 R. radius | 67 R. anterior tibial vein | 94 R. subclavian vein |
| 14 R. vertebral artery | 41 R. ulnar artery I | 68 L. anterior tibial vein | 95 L. subclavian vein |
| 15 R. subclavian artery II | 42 R. interosseous artery | 69 L. posterior tibial vein | 96 R. brachiocephalic vein |
| 16 L. subclavian artery II | 43 R. ulnar artery II | 70 R. popliteal & femoral vein | 97 L. brachiocephalic vein |
| 17 L. vertebral artery | 44 L. ulnar artery I | 71 L. popliteal & femoral vein | 98 Superior vena cava |
| 18 L. internal carotid artery II | 45 L. radius | 72 R. deep femoral vein | 99 Inferior vena cava I |
| 19 L. posterior communicating artery | 46 L. ulnar artery II | 73 L. deep femoral vein | 100 L. renal artery |
| 20 R. posterior communicating artery | 47 L. interosseous artery | 74 R. external & common iliac vein | 101 Abdominal aorta III |

| 21 R. internal carotid artery II | 48 R. deep femoral artery | 75 L. external & common iliac vein | 102 R. renal artery |
| --- | --- | --- | --- |
| 22 Basilar artery | 49 R. femoral artery | 76 Inferior vena cava IV | 103 Abdominal aorta IV |
| 23 L. middle cerebral artery | 50 R. posterior tibial artery | 77 Hepatic vein III | 104 R. renal vein |
| 24 R. middle cerebral artery | 51 R. anterior tibial artery | 78 R. radial vein | 105 Inferior vena cava III |
| 25 L. anterior cerebral artery I | 52 L. femoral artery | 79 R. interosseous vein | 106 L. renal vein |
| 26 R. anterior cerebral artery I | 53 L. deep femoral artery | 80 R. ulnar vein II | 107 Inferior vena cava II |
| 27 L. posterior cerebral artery I | 54 L. anterior tibial artery | 81 R. ulnar vein I | |

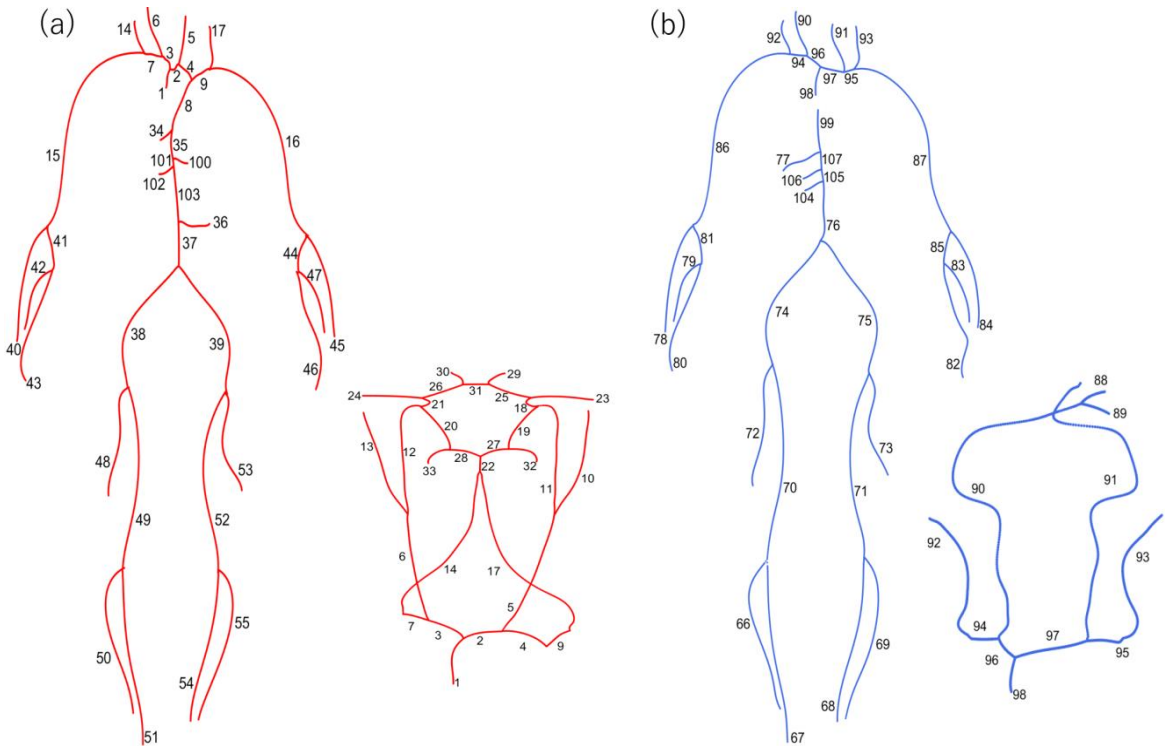

Figure 1. (a) Arterial network comprising 62 arteries, as referred from Formaggia et al. (2009), Müller and Toro (2014) (left). Details of cerebral arteries containing the circle of Willis were adjusted by reference to Alastruey et al. (2007) (right). (b) Venous network comprising 42 veins (left) and details of cerebral veins (right), as referred from Müller et al. (2023) with some adjustments. Numbers refer to those used in Table 1.

*2.1.1. One-dimensional flow model*

The standard one-dimensional (1D) model for blood flow in a vessel is obtainable by averaging the incompressible Navier–Stokes equations over the cross-sectional area $A(x,t)$ under the assumptions that the blood flow is axisymmetric and that it has no swirl (Formaggia et al. 2003). The well-established equations of conservation of mass and momentum can be represented as (Formaggia et al. 2009; Liang et al. 2009):

$$\frac{\partial A}{\partial t}+\frac{\partial Q}{\partial x}=0, \tag{1}$$

$$\frac{\partial Q}{\partial t}+\frac{\partial}{\partial x}\left(\varphi\frac{Q^2}{A}\right)+\frac{A}{\rho}\frac{\partial P}{\partial x}+f=0, \tag{2}$$

where $t$ represents the time and where $x$ denotes the axial coordinate along the vessel. The variables $A(x, t)$, $Q(x, t)$, and $P(x, t)$ respectively denote the cross-sectional area of the vessel, the volume flux, and the average pressure over a cross-section. The momentum correction coefficient $\varphi$ is taken as 1 by assuming a uniform velocity profile. It leads to certain mathematical simplifications (Quarteroni and Formaggia 2004). Here, $f(x, t)$ represents the friction force per unit length; $\rho$ represents the blood density, which is taken as constant.

To complete the governing equation system presented above, a tube law is adopted. It represents the interaction between the internal pressure $P(x, t)$ and the vessel wall displacement, through the cross-sectional area $A(x, t)$ and other parameters, which can be expressed as (Müller and Toro 2014)

$$P(x,t)=P_e+P_0+K\left[\left(\frac{A(x,t)}{A_0(x)}\right)^\alpha-\left(\frac{A(x,t)}{A_0(x)}\right)^\beta\right], \tag{3}$$

where $P_e$ stands for the external pressure, $A_0(x) = A(x, t_0)$ signifies a reference cross-sectional area at a given time point $t_0$, $P_0$ denotes the reference pressure for $A = A_0$, and $K$ is the stiffness coefficient of the material, which depends on the Young's modulus $E_s$ and wall thickness $h$. The specific parameters $\alpha$ and $\beta$ depend on the vessel wall behavior. Throughout this work, referring to earlier studies (Quarteroni and Formaggia 2004; Formaggia et al. 2009; Audebert et al. 2017; Quarteroni et al. 2019) for which full details are given, we consider the values of $K$, $f$, $\alpha$, and $\beta$ for different vascular segments as the following:

$$\text{for an artery}: K=\frac{\pi^{\frac{1}{2}}E_s h}{(1-\sigma^2)A_0^{\frac{1}{2}}}, \quad f=\frac{8\pi\mu Q(x,t)}{\rho A(x,t)}, \quad \alpha=\frac{1}{2}, \quad \beta=0,$$

$$\text{for a vein}: K=\frac{\pi^{\frac{3}{2}}E_s h^3}{12(1-\sigma^2)A_0^{\frac{3}{2}}}, \quad f=\sqrt{\frac{A(x,t)}{A_0(x)}}\frac{8\pi\mu Q(x,t)}{\rho A(x,t)}, \quad \alpha\approx 10, \quad \beta=-\frac{3}{2}, \tag{4}$$

where $\sigma$ is the Poisson ratio, which equals 0.5 for incompressible solids, and $\mu$ represents the blood viscosity.

*2.1.2. Bifurcation and confluence models*

The arterial and venous networks (as shown in Figure 1) with bifurcation and confluence are characterized by the transmission conditions at the interfaces between vessel segments. Mathematical models for bifurcations have been proposed based on many studies (Formaggia et al. 2003, 2009). The developed approach ensures conservation of the mass flux and continuity of the total pressure at the bifurcation, which means that

$$Q_{pa} = Q_{ch_1} + Q_{ch_2}, \tag{5}$$

$$P_{t,pa} = P_{t,ch_1} = P_{t,ch_2}, \tag{6}$$

where *pa* denotes the parent vessel and $ch_1$ and $ch_2$ denote the two child vessels. $P_t = P + \rho/2(Q/A)^2$ represents the total pressure, in which the internal pressure $P(A)$ is calculated using Equation (3). These relations can be complemented with the outflows from the 1D blood vessels at the interface. More details are presented later in Section 3.2 and can be found, for example, in a paper by Formaggia et al. (2009). Considering bifurcation, the following conditions can be imposed in cases of confluence as

$$Q_{pa_1} + Q_{pa_2} = Q_{ch}, \tag{7}$$

$$P_{t,pa_1} = P_{t,pa_2} = P_{t,ch}, \tag{8}$$

where $pa_1$ and $pa_2$ denote the two parent vessels, and *ch* denotes the child vessel.

*2.1.3. Lumped–parameter model*

The micro-vasculatures connect arteries to veins through three compartments in this study, as shown in Figure 2. Each compartment can be modeled as a three-element Windkessel circuit with an inductor, a resistor, and a capacitor, which is also designated as a lumped-parameter model or zero-dimensional (0D) model. In this hydraulic–electric analogy, the blood flow rate and pressure respectively correspond to electric current and voltage/potential (Milišić and Quarteroni 2004). For each compartment, by averaging the 1D model over the length of a vessel under some assumptions, such as neglecting the convective term of the momentum equation, we can obtain (Formaggia et al. 2009; Müller and Toro 2014)

$$\frac{dQ_c}{dt} = \frac{P_c - P_c^{out} - R_c Q_c}{L_c}, \tag{9}$$

$$\frac{dP_c}{dt} = \frac{Q_c^{in} - Q_c}{C_c} + \frac{dP_e}{dt}. \tag{10}$$

Here, variable $Q_c(t)$ stands for the flow rate and $P_c(t)$ denotes the pressure. Also, $R_c$, $L_c$, and $C_c$ respectively represent the resistance, inductance, and capacitance of the lumped compartment. Subscript $c = \{al, cp, vn\}$ respectively denotes arterioles (*al*), capillaries (*cp*), and venules (*vn*). The external pressure $P_e$ is assumed to be zero for this study. $Q_c^{in}(t)$ and $P_c^{out}(t)$ respectively represent the blood inflow rate and blood pressure downstream of the compartment *c*.

In Figure 2, $Q^{in}$ and $Q^{out}$ respectively denote the blood inflow and outflow rates of the micro-vasculature. For the micro-vasculature, we have $Q_{al}^{in} = Q^{in}$ and $Q^{out} = Q_{vn}$.

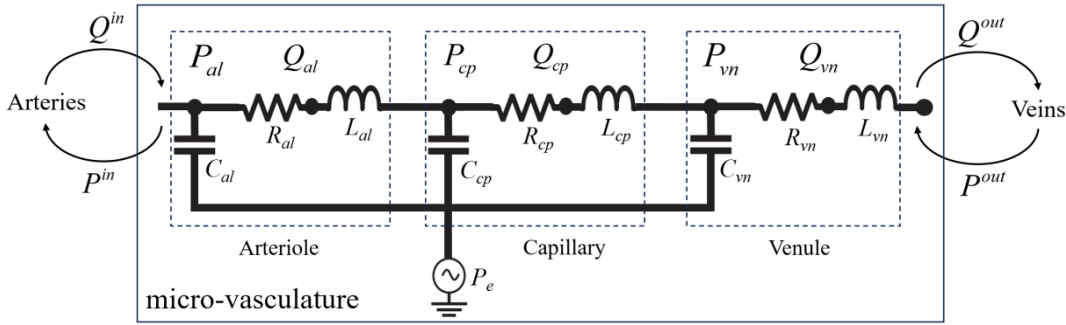

Figure 2. Lumped-parameter network for the micro-vasculature which connects arteries and veins, as taken from Müller and Toro (2014). For each compartment surrounded by a dotted box, we specify compliance *C*, inductance *L*, and resistance *R*, where subscripts denote arterioles (*al*), capillaries (*cp*), and venules (*vn*). $Q^{in}$ represents the blood inflow of the micro-vasculature and is from the arterial outlets. $Q^{out}$ represents the blood outflow of the micro-vasculature and is into the venous inlets. $P^{in}$ and $P^{out}$ respectively represent the pressures upstream and downstream of the micro-vasculature.

Table 2 presents the corresponding connections between arterial and venous networks illustrated in Figure 1. Indexes of linked arteries–veins are the numbers presented in Table 1. A lumped-parameter model is commonly used to describe the simple artery–vein connection, but the subsequent section further extends the use of this model to the specialized compartments representing the organs encompassed within the portal venous network.

Table 2. Connections between arterial and venous networks.
7

| Artery–vein indexes | Upstream artery | Downstream vein |
| --- | --- | --- |
| 32–89 | Left posterior cerebral artery II | Internal cerebral vein |
| 33–89 | Right posterior cerebral artery II | Internal cerebral vein |
| 40–78 | Right radial artery | Right radial vein |
| 100–106 | Left renal artery | Left renal vein |
| 102–104 | Right renal artery | Right renal vein |
| 42–79 | Right interosseous artery | Right interosseous vein |
| 43–80 | Right ulnar artery II | Right ulnar vein II |
| 45–84 | Left radial artery | Left radial vein |
| 46–82 | Left ulnar artery II | Left ulnar vein II |
| 47–83 | Left interosseous artery | Left interosseous vein |
| 10–93 | Left external carotid artery | Left external jugular vein |
| 13–92 | Right external carotid artery | Right external jugular vein |
| 23–91 | Left middle cerebral artery | Superior sagittal sinus |
| 24–88 | Right middle cerebral artery | Inferior sagittal sinus |
| 29–91 | Left anterior cerebral artery II | Left internal jugular vein |
| 30–88 | Right anterior cerebral artery II | Inferior sagittal sinus |
| 48–72 | Right deep femoral artery | Right deep femoral vein |
| 50–66 | Right posterior tibial artery | Right posterior tibial vein |
| 51–67 | Right anterior tibial artery | Right anterior tibial vein |
| 53–73 | Left deep femoral artery | Left deep femoral vein |
| 54–68 | Left anterior tibial artery | Left anterior tibial vein |
| 55–69 | Left posterior tibial artery | Left posterior tibial vein |

## *2.2. Portal venous network*

Various components of the portal venous system and related organs are presented in Figure 3, including the intestine, stomach, spleen, and liver, as well as the corresponding arteries, veins, and portal veins. It is noteworthy that substances which play key roles in maintaining our lives, such as glucose and insulin, are secreted or absorbed in this system. Glucose absorbed by the intestine can be supplied through portal veins (Nos. 59, 61, 63). Insulin produced by pancreatic β cells in the pancreas is secreted into the portal vein (No. 60) and is subsequently delivered into the liver (Koh et al. 2022) through portal veins (Nos. 61, 63). After the liver extracts part of the insulin, the remaining part is delivered to the systemic circulation via the hepatic vein (No. 65).

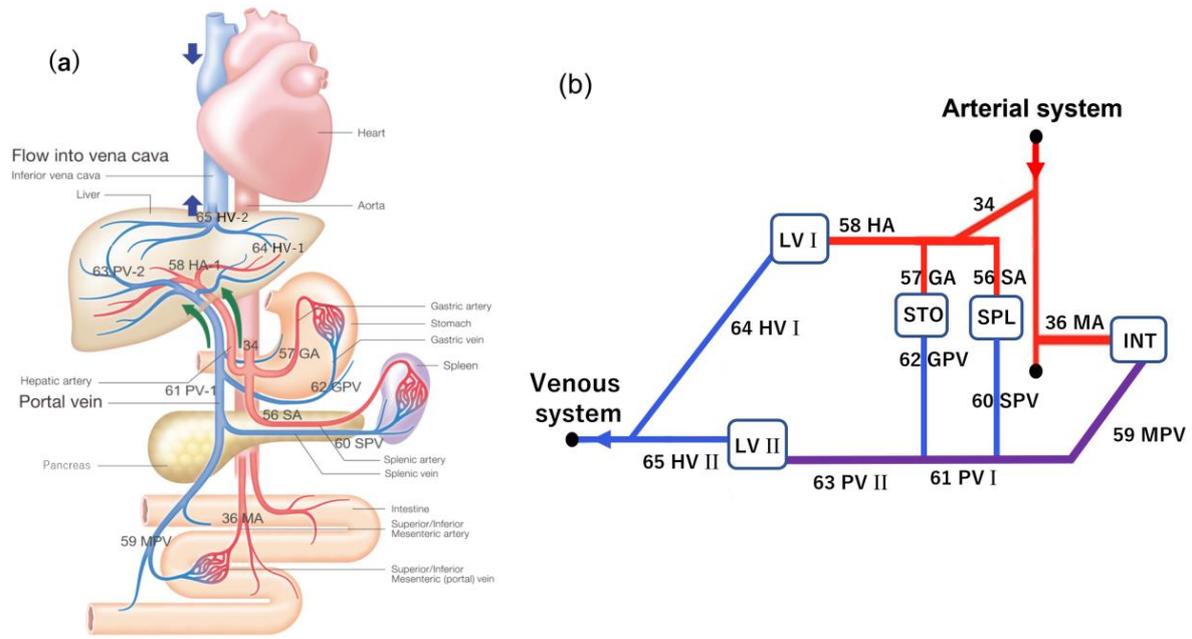

Figure 3. (a) Portal venous system and related organs. For numerical simulations, we consider six arteries, four veins, and three portal veins, which provide pathways for blood and substances transported between organs: stomach, spleen, pancreas, intestine, and liver. (b) A schematic representation of the portal venous system corresponding to panel (a). Lines represent the arteries (red), veins (blue), and portal veins (purple), which are described using 1D models and thereby differ from the electric analog circuits presented in Audebert et al. (2017). The abbreviation for uppercase letters surrounded by a blue box represents the lumped-parameter model for the micro-vasculature of organs, as depicted in Table 3. Black circles represent the junctions to other arteries and veins. Numbers of segments refer to those used in Table 1.

Blood flow in vessels and organs is represented respectively using 1D and 0D models, as explained in Section 2.1. Table 3 provides lumped-parameter models used for organs, along with their corresponding upstream and downstream segments and their respective abbreviations. Considering the structural complexity of the liver, blood flow from the hepatic artery and portal vein enters the veins within the liver through two compartments, as illustrated in Figure 3(b).

Table 3. Lumped-parameter models for organs.

| Lumped model | Organ | Upstream artery | Abbreviation | Downstream vein | Abbreviation |
| --- | --- | --- | --- | --- | --- |
| SPL | Spleen | Splenic artery | SA | Splenic portal vein | SPV |
| STO | Stomach | Gastric artery | GA | Gastric portal vein | GPV |
| INT | Intestine | Mesenteric artery | MA | Mesenteric portal vein | MPV |
| LV I | Liver | Hepatic artery | HA | Hepatic vein I | HV I |
| LV II | Liver | Hepatic portal vein II | PV II | Hepatic vein II | HV II |

From Figure 3(b), a different type of junction is noted as existing at the celiac artery (No. 34), which is branched into three segments connecting three organs. For such transmission conditions, conservation of the mass flux and the continuity of the total pressure at the bifurcation should be expressed as

$$Q_{pa} = Q_{ch_1} + Q_{ch_2} + Q_{ch_3}, \tag{11}$$

$$P_{t,pa} = P_{t,ch_1} = P_{t,ch_2} = P_{t,ch_3}, \tag{12}$$

where *pa* denotes the parent vessel and where $ch_1$, $ch_2$, and $ch_3$ respectively denote the corresponding three child vessels.

### *2.3. Heart–pulmonary circulation*

Taking $Q_{av}(t)$ as representing the flow rate through the aortic valve as an example, its time variation can be described as

$$\frac{dQ_{av}}{dt} = \frac{P_{lv} - P_{ao} - R_{av}Q_{av} - B_{av}Q_{av}|Q_{av}|}{L_{av}} \times D_{av}(P_{lv}, P_{ao}) \tag{13}$$

with

$$D_{av}(P_{lv}, P_{ao}) = \begin{cases} 1 & P_{lv} > P_{ao}, \\ 0 & P_{lv} \leq P_{ao}. \end{cases}$$

Here $L_{av}$ and $R_{av}$ respectively signify the coefficients for inertial terms and viscous losses. $B_{av}$ is the Bernoulli resistance of the valves, which is associated with convective acceleration and dynamic pressure losses. Cardiac valve function $D_{av}$ is defined as an ideal diode model, which opens and closes instantaneously depending on the changing of the pressure gradient sign between the left ventricle ($P_{lv}$) and aorta ($P_{ao}$). Blood flow from the left ventricle to the aorta is represented by the variable $Q_{av}$. Moreover, by imposing mass conservation, mass fluxes from the venous network are used to update the volume of right atrium ($V_{ra}$) as

$$\frac{dV_{ra}}{dt} = Q_{vc} - Q_{tv}, \tag{14}$$

where $Q_{vc}$ and $Q_{tv}$ respectively represent the flow from the vena cava and through the tricuspid valve. With Equations (13) and (14), the heart–pulmonary circulation can be connected to the arterial and venous trees, respectively, via variables $Q_{av}$ and $Q_{vc}$.

The time-varying elastance method (Sun et al. 1997; Formaggia et al. 2009) is the basis of mathematical model for calculating the pressure ($P_{cm}$) in the four cardiac chambers: the right atrium (*ra*), right ventricle (*rv*), left atrium (*la*), and left ventricle (*lv*). The calculation is

$$P_{cm}(t) = (E_{cm,A}e(t) + E_{cm,B})(V_{cm} - V_{cm,0}) + S_v \frac{dV_{cm}}{dt}, \quad cm = \{ra,\ rv,\ la,\ lv\}, \tag{15}$$

where $E_{cm,A}$ and $E_{cm,B}$ respectively represent the active and passive elastances of the four cardiac chambers. In addition, $V_{cm}$ stands for the current chamber volume. $V_{cm,0}$ denotes the dead volume, assumed here as 0. $S_v$ expresses the viscoelasticity coefficient of cardiac wall. Finally, $e(t)$ signifies the normalized time-varying elastance, which for the atria is

$$e_a(t) = \begin{cases} \frac{1}{2}\{1 + \cos[\pi(t + \tau - t_{ar})/\tau_{arp}]\} & 0 \le t \le t_{ar} + \tau_{arp} - \tau, \\ 0 & t_{ar} + \tau_{arp} - \tau < t \le t_{ac}, \\ \frac{1}{2}\{1 - \cos[\pi(t - t_{ac})/\tau_{acp}]\} & t_{ac} < t \le t_{ac} + \tau_{acp}, \\ \frac{1}{2}\{1 + \cos[\pi(t - t_{ar})/\tau_{arp}]\} & t_{ac} + \tau_{acp} < t \le \tau, \end{cases} \tag{16}$$

and for ventricles is

$$e_v(t) = \begin{cases} \frac{1}{2}[1 - \cos(\pi t / \tau_{vcp})] & 0 \le t \le \tau_{vcp}, \\ \frac{1}{2}\{1 + \cos[\pi(t - \tau_{vcp})/\tau_{vrp}]\} & \tau_{vcp} < t \le \tau_{vcp} + \tau_{vrp}, \\ 0 & \tau_{vcp} + \tau_{vrp} < t \le \tau. \end{cases} \tag{17}$$

Here, $\tau$ represents the cardiac cycle duration. Furthermore, $\tau_{acp}$, $\tau_{vcp}$, $\tau_{arp}$, and $\tau_{vrp}$ respectively represent the duration of atrial/ventricular contraction/relaxation. Also, $T_{ac}$ and $t_{ar}$ respectively denote the time points in one cardiac cycle at which the atria begin to contract and relax.

Detailed discussion of heart–pulmonary models can be found in papers reported by Sun et al. (1997), Liang et al. (2009), and Quarteroni et al. (2017).

## 3. Numerical procedure for the entire 1D-0D closed-loop network

Numerical schemes for 1D model are combined through the transmission conditions defined for bifurcation/confluence junctions and through 0D model for the artery–vein connection as follows.

### *3.1. Numerical schemes for 0D and 1D models*

Fourth-order Runge–Kutta method is adopted to solve 0D models in this study. For 1D models, Equations (1)–(3) can be transformed into algebraic form. They are solvable using the two-step Lax–Wendroff scheme, which has second-order accuracy in space and time. A rectilinear grid is defined for $\{0 \leq x \leq L\} \times \{t \geq 0\}$ with grid spacing $\Delta x > 0$ and $\Delta t > 0$, respectively, in the space and time coordinate directions, where $\Delta x = L/I$ with $I$ being a positive integer. Let $A^n_i = A(x_i, t_n)$ and $Q^n_i = Q(x_i, t_n)$ at the grid point $(x_i, t_n) = (i\Delta x, n\Delta t)$.

For the interior points, $i \in [1, I-1]$, the cross-sectional area $A(x, t)$ and flow $Q(x, t)$ at the intermediate time-step $n + 1/2$ are calculable as

$$A_j^{n+\frac{1}{2}} = \frac{1}{2}\left(A^n_{j+\frac{1}{2}} + A^n_{j-\frac{1}{2}}\right) - \frac{\Delta t}{2\Delta x}\left(Q^n_{j+\frac{1}{2}} - Q^n_{j-\frac{1}{2}}\right),$$
$$Q_j^{n+\frac{1}{2}} = \frac{1}{2}\left(Q^n_{j+\frac{1}{2}} + Q^n_{j-\frac{1}{2}}\right) - \frac{\Delta t}{2\Delta x}\left(G^n_{j+\frac{1}{2}} - G^n_{j-\frac{1}{2}}\right) + \frac{\Delta t}{4}\left(G^n_{F_{j+\frac{1}{2}}} + G^n_{F_{j-\frac{1}{2}}}\right),$$
(18)

where

$$j = i + \frac{1}{2} \quad \text{or} \quad i - \frac{1}{2}, \qquad G = \frac{Q^2}{A} + \frac{K}{\rho}\left(\frac{\alpha}{\alpha+1}\frac{A^{\alpha+1}}{A_0^{\alpha}} - \frac{\beta}{\beta+1}\frac{A^{\beta+1}}{A_0^{\beta}}\right), \qquad G_F = -f, \quad (19)$$

with α, β, K, and $f$ provided in Equation (4) for artery and vein values. Then, $A$ and $Q$ at the new time-step are calculated as

$$A_i^{n+1} = A_i^n - \frac{\Delta t}{\Delta x}\left(Q_{i+\frac{1}{2}}^{n+\frac{1}{2}} - Q_{i-\frac{1}{2}}^{n+\frac{1}{2}}\right),$$
$$Q_i^{n+1} = Q_i^n - \frac{\Delta t}{\Delta x}\left(G_{i+\frac{1}{2}}^{n+\frac{1}{2}} - G_{i-\frac{1}{2}}^{n+\frac{1}{2}}\right) + \frac{\Delta t}{2}\left(G_{F_{i+\frac{1}{2}}}^{n+\frac{1}{2}} + G_{F_{i-\frac{1}{2}}}^{n+\frac{1}{2}}\right).$$
(20)

### *3.2. Junctions between 1D segments*

Mathematical analysis of such a system emphasizes its hyperbolic nature (Formaggia et al. 2003). By introducing the so-called Riemann invariants, variations of blood flow $Q$ and cross-sectional area $A$ can be treated as combinations of forward-wave $W_f$ and backward-wave $W_b$, as

$$W_{f,b} = \frac{Q}{A} \pm 4(c - c_0) \tag{21}$$

with

$$c = \sqrt{\frac{A}{\rho}\frac{\partial P}{\partial A}} = \sqrt{\frac{K}{\rho}\left[\alpha\left(\frac{A}{A_0}\right)^\alpha - \beta\left(\frac{A}{A_0}\right)^\beta\right]}, \tag{22}$$

which represents the speed of pulse waves relative to blood flow and $c_0 = c(A_0)$.

The system of Equations (5) and (6) describing the transmission conditions is taken as an example. Six unknowns ($A_{pa}$, $Q_{pa}$), ($A_{ch1}$, $Q_{ch1}$), and ($A_{ch2}$, $Q_{ch2}$), which depend only on time $t$, should be considered at the bifurcation. Three other conditions are necessary for solving it. Here, we assume that the interface conditions account only for entering waves (Formaggia et al. 2006; Quarteroni and Valli 2008). The effect of the parent vessel on the bifurcation/confluence appears on the forward wave. The effect of the child vessels is apparent on the backward waves. Then the transmission conditions at each time step $t^{n+1}$ can be deduced as

$$\begin{aligned}
\frac{Q_{pa}^{n+1}}{A_{pa}^{n+1}} + 4\left(c\left(A_{pa}^{n+1}\right) - c\left(A_{0,pa}\right)\right) &= W_f^{n+1}, \\
\frac{Q_{ch_1}^{n+1}}{A_{ch_1}^{n+1}} - 4\left(c\left(A_{ch_1}^{n+1}\right) - c\left(A_{0,ch_1}\right)\right) &= W_{b_1}^{n+1}, \\
\frac{Q_{d_2}^{n+1}}{A_{ch_2}^{n+1}} - 4\left(c\left(A_{ch_2}^{n+1}\right) - c\left(A_{0,ch_2}\right)\right) &= W_{b_2}^{n+1}.
\end{aligned} \tag{23}$$

Referring to Formaggia et al. (2003), we adopt the first-order approximation of the outgoing characteristic variables at the boundary points $x = 0, L$ as

$$\begin{aligned}
W_f^{n+1} &= W_f(t^n, L - \lambda_f^n(L)\Delta t), \\
W_{b_1}^{n+1} &= W_b(t^n, -\lambda_{b_1}^n(0)\Delta t), \\
W_{b_2}^{n+1} &= W_b(t^n, -\lambda_{b_2}^n(0)\Delta t).
\end{aligned} \tag{24}$$

Therein, $\lambda_{f,b} = Q/A \pm c$. The values of $W_f$ and $W_b$ is calculable by interpolating between the values at the two nearest spatial nodes of the associated parent and child vessels. Furthermore, for the child vessels, $W_b = 0$ might be set as a perfectly non-reflecting boundary condition to force a null backward wave (Formaggia et al. 2006). With Equation (24), the system of six nonlinear equations (5), (6), and (23) for six variables $\{Q_{pa}^{n+1}, A_{pa}^{n+1}, Q_{ch_1}^{n+1}, A_{ch_1}^{n+1}, Q_{ch_2}^{n+1}, A_{ch_2}^{n+1}\}$ is solvable using Newton–Raphson procedure. A more detailed discussion of this numerical implementation can be found in the article of an earlier study by Sherwin et al. (2003).

### *3.3. Minimization problem for a closed-loop model*

As shown in Figures 1 and 3, we have included 107 blood vessels, 27 micro-vasculatures, and a heart–pulmonary circulation in our closed-loop model for the blood circulation network which consists of four subsystems: 1D models (designated as *F*); 0D lumped-parameter models (designated as *C*); a heart–pulmonary model (designated as *H*); along with 53 junctions between 1D segments described by bifurcation and confluence models (designated as *R*). Based on the four subsystems, an iterative algorithm for the duration of a cardiac cycle $\tau$ is designed.

For any iteration $k = 1, 2, \ldots$, we have

$$
\begin{aligned}
(Q_{\mathbf{N}_F}^k, P_{\mathbf{N}_F}^k) &= F(Q_{\mathbf{N}_F}^{k-1}, P_{\mathbf{N}_F}^{k-1}, Q_{\mathbf{N}_C}^{k-1}, P_{\mathbf{N}_C}^{k-1}, Q_{\mathbf{N}_H}^{k-1}, P_{\mathbf{N}_H}^{k-1}, Q_{\mathbf{N}_R}^{k-1}, P_{\mathbf{N}_R}^{k-1}), \\
(Q_{\mathbf{N}_C}^k, P_{\mathbf{N}_C}^k) &= C(Q_{\mathbf{N}_F}^{k-1}, P_{\mathbf{N}_F}^{k-1}), \\
(Q_{\mathbf{N}_H}^k, P_{\mathbf{N}_H}^k) &= H(Q_{\mathbf{N}_F}^{k-1}, P_{\mathbf{N}_F}^{k-1}), \\
(Q_{\mathbf{N}_R}^k, P_{\mathbf{N}_R}^k) &= R(Q_{\mathbf{N}_F}^{k-1}, P_{\mathbf{N}_F}^{k-1}).
\end{aligned}
\tag{25}
$$

Here, $\mathbf{N}_F$, $\mathbf{N}_C$, $\mathbf{N}_H$, and $\mathbf{N}_R$ are the sets of nodes for which the values are computed respectively by subsystems *F*, *C*, *H*, and *R*. Nodes included in $\mathbf{N}_F$ are solved using a 1D model with the values from 0D lumped-parameter models, or the heart–pulmonary model, or bifurcation and confluence models connected upstream and downstream of the 1D model computed in the previous iteration step. Nodes included in $\mathbf{N}_C \cup \mathbf{N}_H \cup \mathbf{N}_R$ are solved using the values by both upstream and downstream 1D models computed in the previous iteration step.

Objective function J, which represents the discrepancy between two consecutive iterations during a cardiac cycle $\tau$, is defined as

$$
\mathbf{J}_k = \frac{\sum_{\omega \in \mathbf{N}} \dfrac{\max\limits_{t \in [0,\, \tau]} |Q_\omega^k - Q_\omega^{k-1}|}{\max\limits_{t \in [0,\, \tau]} |Q_\omega^{k-1}|} + \dfrac{\max\limits_{t \in [0,\, \tau]} |P_\omega^k - P_\omega^{k-1}|}{\max\limits_{t \in [0,\, \tau]} |P_\omega^{k-1}|}}{\text{num}(\mathbf{N})},
\tag{26}
$$

where $\mathbf{N} \equiv \mathbf{N}_F \cup \mathbf{N}_C \cup \mathbf{N}_H \cup \mathbf{N}_R$ and where num($\mathbf{N}$) represents the number of elements of set $\mathbf{N}$.

*3.4. Parameters for models*

For the global closed-loop vascular network considered in this study, all the geometrical parameters and pulse wave velocities $c$ used to estimate the corresponding elastic parameters $E_s$, are taken from works reported by Müller and Toro (2014) and Toro et al. (2022). The values of parameters related to the portal venous system shown in Figure 3 refer to the studies conducted by Mynard and Smolich (2015) and by Audebert et al. (2017). Additionally, parameters for the heart–pulmonary circulation are from papers by Sun et al. (1997) and by Formaggia et al. (2009).

Given these parameters, a thorough simulation of the closed-loop blood circulation can be implemented. The 1D domains are divided into grids with a length of $\Delta x = 0.1$ cm to ensure, at least, ten computational grids in each vessel segment with the time step $\Delta t = 5 \times 10^{-4}$ s. The duration of a cardiac cycle $\tau$ is set to 1.0 s. The friction function is defined as given in Equation (4), with blood viscosity $\mu = 0.0045$ Pa·s and blood density $\rho = 1050$ kg/m$^3$. In this study, $Q(x, 0) = 0$ and $A(x, 0) = A_0$ are assigned as initial conditions.

## 4. Results

We used a CPU with 24 cores (Xeon Silver 4214R; Intel Corp.), with 64GB RAM. In our computation procedure, the presented model typically converges to a periodic state over approximately 30 iterations. Figure 4 displays the convergence process of the objective function with iterations until reaching the target value of 0.05, which serves as the stopping criterion for the iteration process.

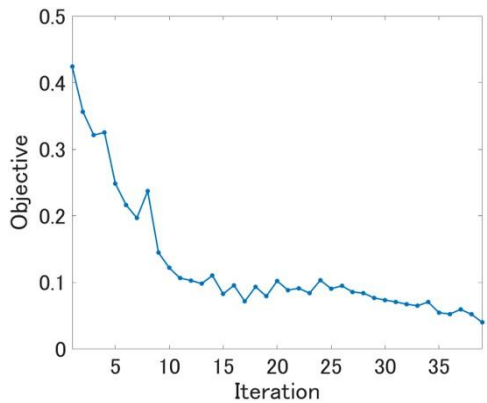

Figure 4. Objective function J$_k$ versus iterations $k$, as depicted in Equation (26).

*4.1. Numerical results*

Figure 5 displays the computed blood flow $Q(x, t)$ at different time points of a cardiac cycle in the entire circulation system, including arterial, venous, and portal venous networks. In it, the vessel line thickness represents the magnitude of the flow at the locations. From Figure 5, one can observe the following.

- In the arterial system, as time $t$ increases, the blood flows out from the heart through the aorta; then it is gradually delivered to the entire body. In our simulation, the pulse wave transit time (PWTT) is calculated as the time difference of the pulse wave from the aorta to arrival at different locations. The PWTT is almost determined by the distance from the aorta and the pulse wave velocity (PWV), which depends on the elastance of blood vessels. In this simulation, the maximum value of the calculated PWTTs for arteries is 0.29 s. Additionally, the corresponding average PWV is 569.85 cm/s. Both findings are within their respective physiological ranges (Barral and Croibier 2011).
- In the portal venous system, at time point $t = 0.16$ s, blood mainly comes from the stomach, spleen, and intestine, causing the volumes of the capillaries in these organs to decrease because of the difference of blood flows $(Q^{in}-Q^{out})<0$. At time points $t = 0.20$ s and 0.25 s, their volumes are increased by blood supply from the arteries. A difference exists between the time points at which the volumes of capillaries begin to increase in the intestine and other organs because of the difference of their distances from the heart, which are visible from Figure 5(a) at the same time points ($t = 0.20$ s and 0.25 s). As time progresses from 0.31 s to 1.1 s in Figure 5(b), the portal venous system returns to its supplement state at time point $t = 0.16$ s.
- In the venous system, unlike the arterial and portal venous systems, distinguishing changes throughout the cycle are not seen based on the physiological characteristics of more or less stationary blood flow in the venous system.

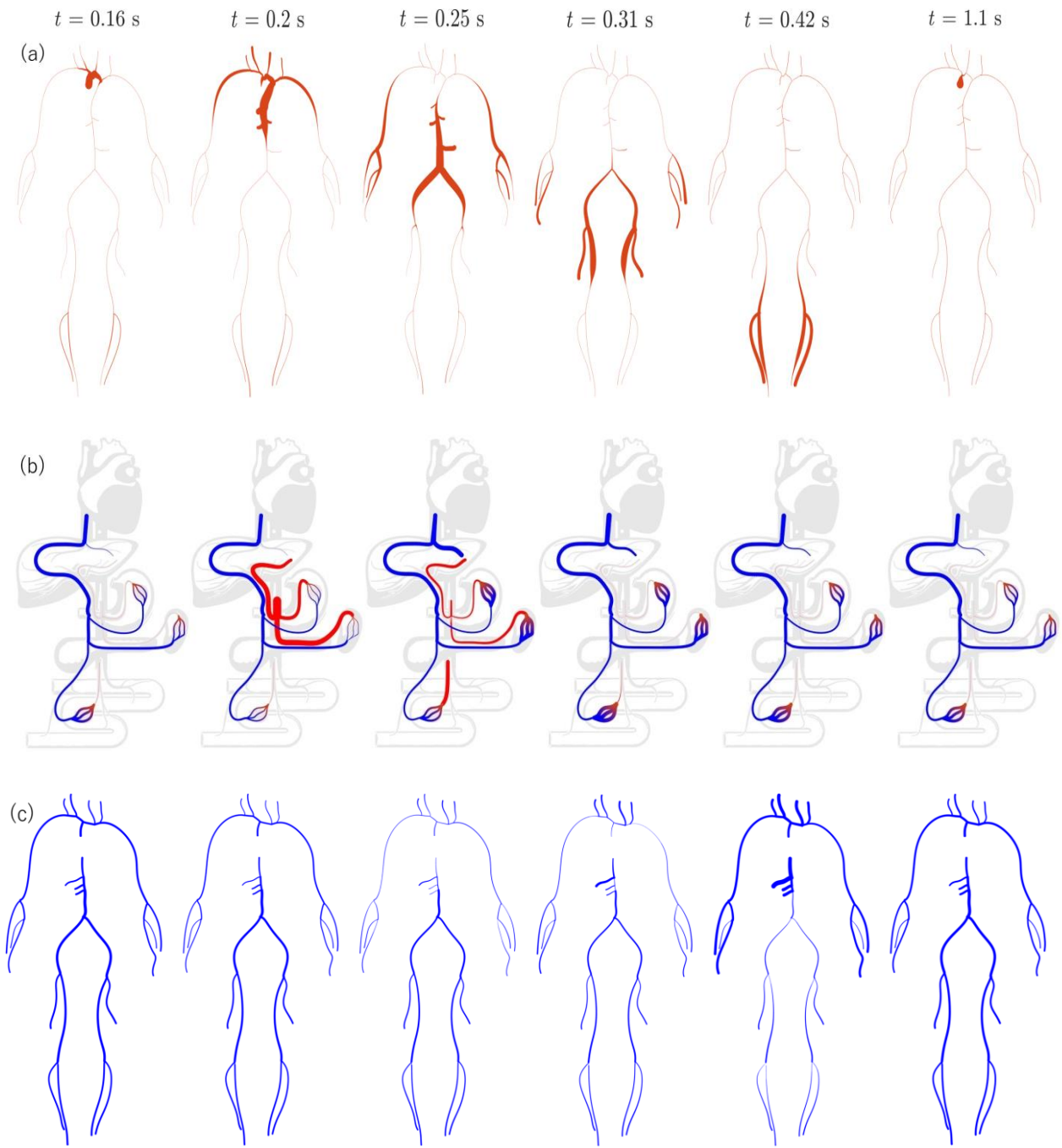

Figure 5. Schematic representations of the computed blood flows along the arterial (a), portal venous (b), and venous (c) networks throughout one cardiac cycle (1 s). Red lines represent arteries. Blue lines represent veins and portal veins. In (a) and (b), the line thickness at a spatial–temporal position ($x$, $t$) is determined by the computed value of blood flow $Q(x, t)$. In (c), the line thickness is adjusted relative to $Q(x,t)/\max_{t\in[0, 1]}|Q(x,t)|$ for capturing the minor changes at each segment in the venous network during the cycle.

## 4.2. Validation

Although the closed-loop model, which consists of complex multiscale models, makes it possible to study widely diverse physiological and pathological conditions, its in-vivo validation is quite difficult because of the complexity. Another difficulty arises from the fact that a fully quantitative validation would require a subject-specific approach, in which all parameters defining the multiscale model (including geometries, viscoelastic properties, peripheral impedances, and varying elastance of the heart) must be measured or estimated precisely for a specific subject. Therefore, hereinafter, we present some simulations and compare them with physiological data reported in the literature to observe whether they match the general physiological data or not.

*4.2.1. Pressures and volumes in cardiac chambers*

For heart–pulmonary circulation, Figure 6 shows the computed pressures and volumes for the four cardiac chambers, where pressure and volume variations during the cardiac cycle agree well with the physiological conditions described by Levick (2013) (whether atria or ventricle).

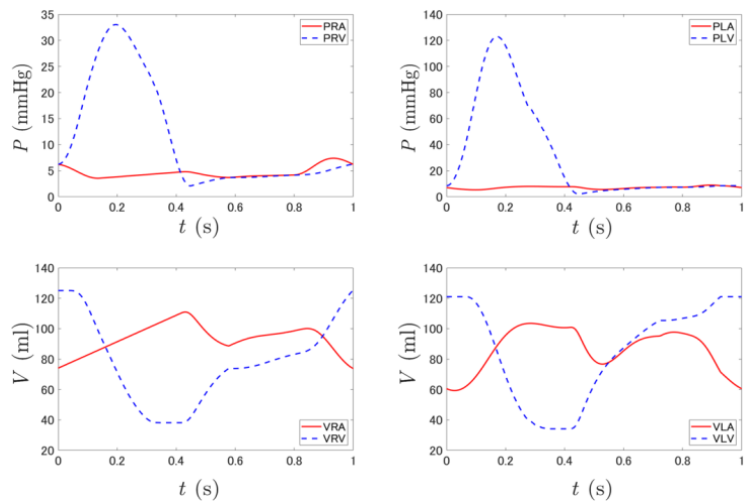

Figure 6. Computed pressures $P$ and volumes $V$ for the four cardiac chambers: right atrium (RA), right ventricle (RV), left atrium (LA), and left ventricle (LV).

*4.2.2. Flows in blood vessels*

Figures 7 and 8 present the computed waveforms of flow rates at the midpoints of selected vessel segments in the arterial, venous, and portal venous networks. In Figure 7, the simulated waveforms at various segments of the systemic arteries are compared to those from the global model provided by Müller and Toro (2014). The comparisons exhibit a satisfactory level of agreement. Furthermore, one might observe that, as the blood flows away from heart towards the periphery, the peak flow decreases progressively, which well reproduces physiological patterns.

The bar chart in Figure 9 portrays a comprehensive comparison of the simulation results derived from the present work with results computed by Müller and Toro (2014) and Mynard and Smolich (2015), and with experimentally obtained data (Zitnik et al. 1965; Murgo et al. 1980; Wolf et al. 1993; Cheng et al. 2003; Gorg et al. 2002; Barakat 2004). The comparison reveals that flow rates simulated in this study at different segments of the systemic arteries, veins, and portal veins, exhibit a satisfactory level of agreement with the data reported in the literature, and are also within the physiological ranges.

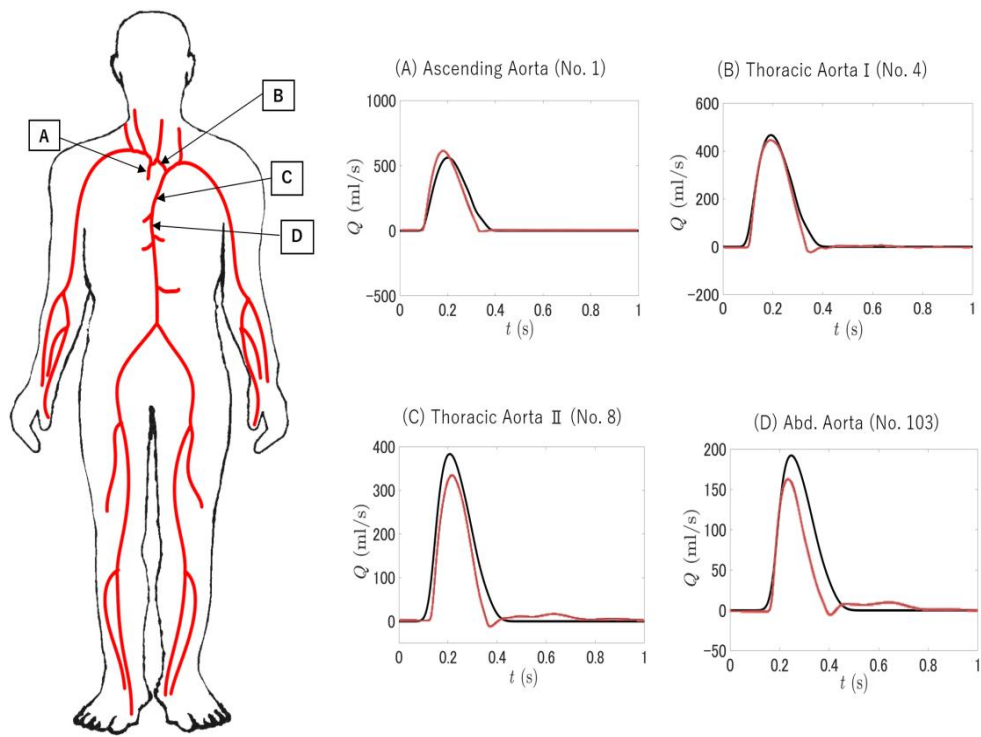

Figure 7. Computed waveforms (black lines) and results obtained using the Müller–Toro model (Müller and Toro 2014) (red lines) along the aorta.

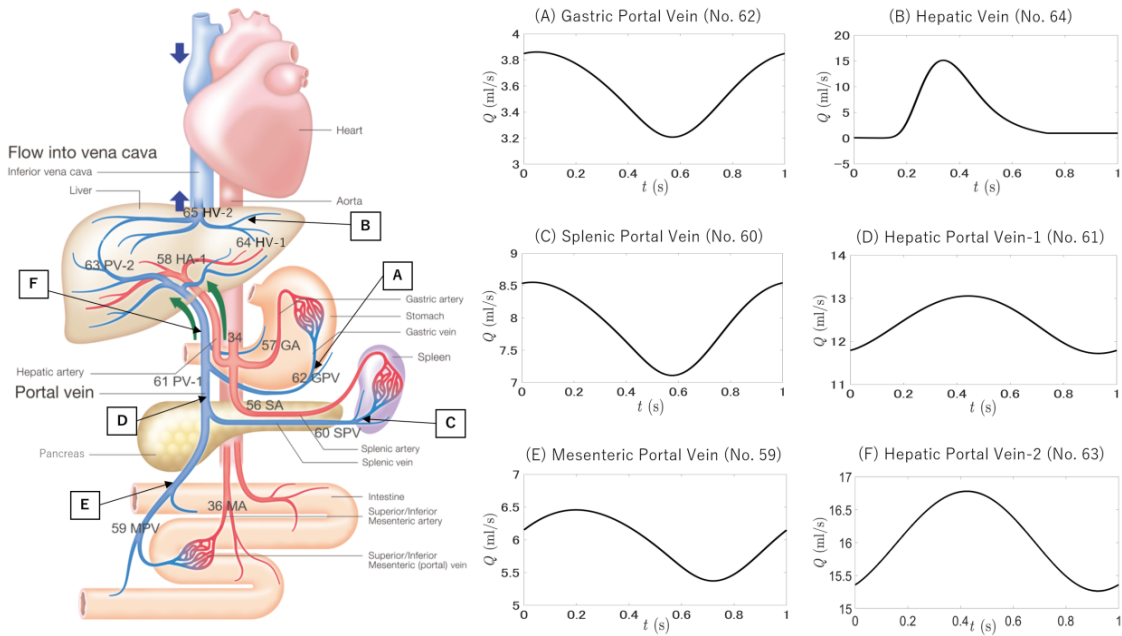

Figure 8. Computed waveforms along the portal venous system.

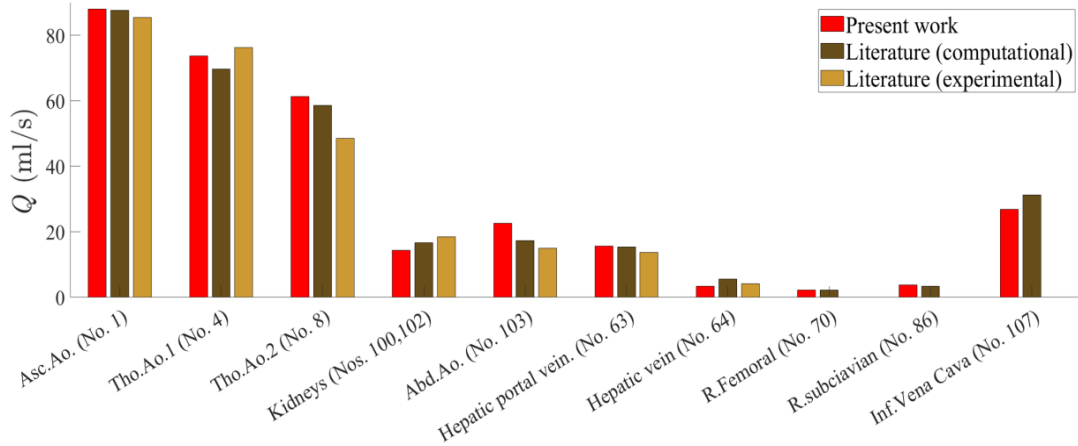

Figure 9. Blood flow in selected systemic arteries, veins, and portal veins: computational results of the present model; computational results from Müller and Toro (2014) (Nos. 1, 4, 8, 103, 100, 102, 70, 86, and 107) and Mynard and Smolich (2015) (Nos. 63 and 64); and data from Murgo et al. (1980) for ascending aorta (No. 1), from Zitnick et al. (1965) for thoracic aorta 1 (No. 4), from Cheng et al. (2003) for thoracic aorta 2 (No. 8) and abdominal aorta (No. 103), from Wolf et al. (1993) for kidneys, from Gorg et al. (2002) for Hepatic portal vein (No. 63), and from Barakat (2004) for Hepatic vein (No. 64).

## 5. Parallel computation

Numerical simulation of the constructed closed-loop network faces a daunting computing challenge because the execution time taken for the computation of 30 iterations, which is necessary to reach the periodic state as described in Section 3.4, exceeds 170 hr in our personal computer environment with 24 cores. Such time-consuming computations severely restrict the practical application of this global multiscale model even if we were to use supercomputers, prompting us to explore parallelization strategies appropriate to our specific problem.

## 5.1. Parallelization performance

In this context, speedup ratio $S$ and parallel-efficiency $E$ are

$$S(p) = \frac{T(1)}{T(p)} \quad \text{and} \quad E(p) = \frac{S(p)}{p}.$$

They are introduced to evaluate the parallel computations achieved by Equations (25) and (26) when using multiple cores. Here $T(p)$ denotes the execution time when employing $p$ cores; $T(1)$ represents the time taken in serial computation. Figure 10 presents a comparison of the speedup $S$ and parallel-efficiency $E$ obtained by varying the number of cores involved in the processing.

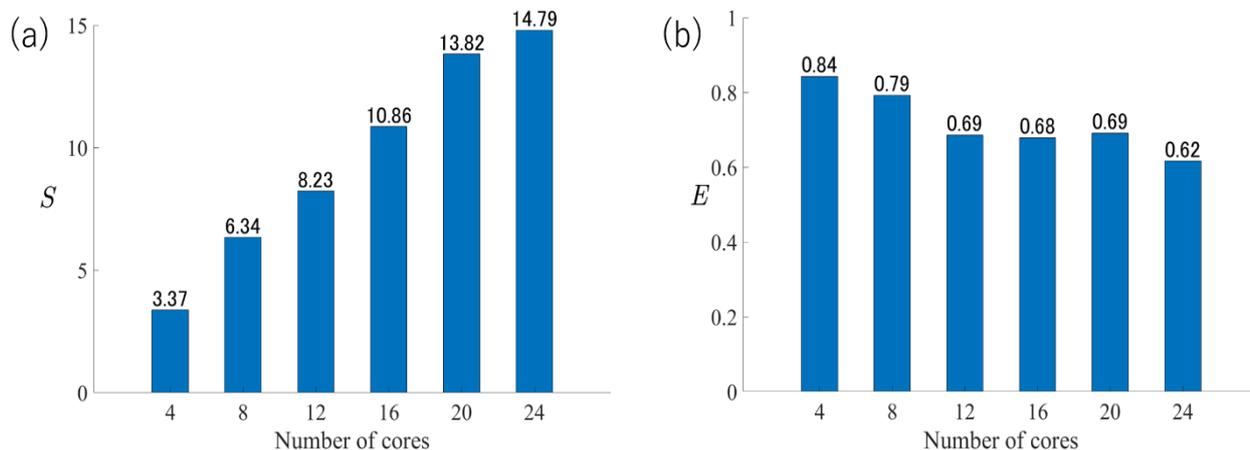

Figure 10. (a) Speedup $S$ and (b) parallel-efficiency $E$ against the number of cores.

According to Figure 10(a), the speedup ratio $S$ increases almost linearly as the number of cores increases. However, the parallel-efficiency $E$ decreases with multiple cores, as is often observed in parallel computations.

Table 4. Average execution times for subsystems using 24 cores.

| Subsystem | $F$ | $R$ | $C$ | $H$ |
|---|---|---|---|---|
| Time (s) | 3.847 | 422.218 | 0.159 | 1.644 |

Table 4 presents that subsystem $R$, which is solved using the iterative Newton–Raphson method, has a markedly higher time cost than other subsystems $F$, $C$, and $H$ defined in Equation (25). In such cases, parallel task scheduling, also designated as parallel job scheduling, can be employed to improve performances in terms of parallel program efficiency and load balance. To achieve this, we use the Shortest Job First (SJF) scheduling strategy, which assigns priority to the job with the shortest (expected) execution time (Vignesh et al. 2013). In addition to SJF scheduling, we further consider partitioning a data structure into substructures, which is a frequently used operation for the extraction of data parallelism (Banerjee and Browne 1996).

*5.2. Parallelization with substructures*

In this subsection, computation substructures are introduced to divide tasks with long execution times into smaller and parallelizable sub-tasks. During the computation process, for subsystem $R$, which represents bifurcation and confluence models, it is noteworthy that the iterative scheme is computed independently for each time point $t^n \in [0, \tau]$. The computation at each time point relies on the corresponding transmission condition, which is determined by the blood flow from relevant blood vessels, rather than being influenced by other time points within the interval $[0, \tau]$, which indicates that the most time-consuming iterative computation for subsystem $R$, as shown in Table 4, is executed independently at each time point. Taking ideas from the partitioning strategy with domain decomposition developed by Berger and Bokhari (1987), we partition the interval $[0, \tau]$ into smaller sub-intervals for construction of the data substructures. The computational work of the iteration can thereby be decomposed into sub-tasks executed on these data substructures. Such parallelizable sub-tasks would be distributed over multicores, allowing them to be processed simultaneously. This approach, denoted as interval partitioning herein, is useful for parallel computation. Although the total number of tasks increases, interval partitioning works on reducing the execution time of divided sub-tasks, and ultimately decreasing the overall execution time by balancing the workload across multiple cores.

Figure 11 presents the marked performance degradation which might occur with multicore systems when the number of tasks with long execution times does not match the number of available cores, resulting in idle cores. The tasks, represented by blue rectangles in Figure 11, are units of work. Each of them is performed on one core, without communication between different cores. Therefore, partitioning such a task with the time domain $[0, \tau]$ into sub-tasks with the corresponding time domains of $[0, \tau/2]$ and $[\tau/2, \tau]$, which can then be assigned to multiple cores, is a suitable approach to resolve such an issue of idle cores. This approach has the potential to increase core utilization and to decrease the total computation time.

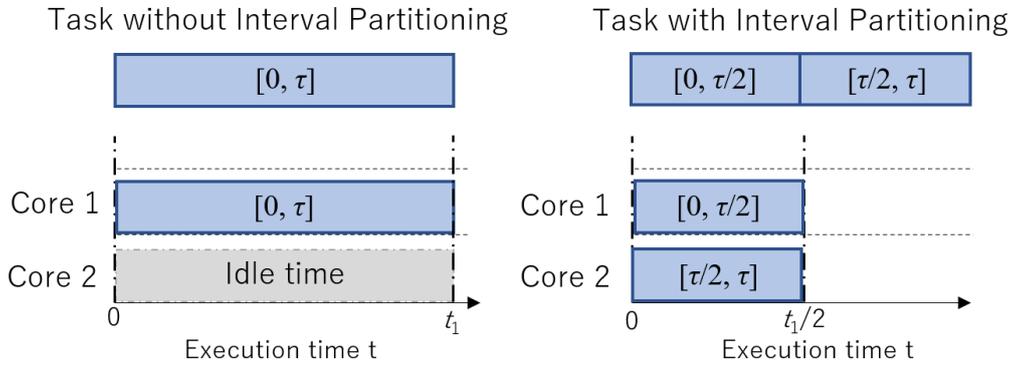

Figure 11. Example task scheduling with and without interval partitioning.

Consequently, in addition to the commonly used SJF scheduling, this study explores the interval partitioning strategy for executing tasks involving subsystem $R$. We denote an algorithm as M$k$, with $k$ representing the number of sub-tasks into which a task for subsystem $R$ with a long execution time is partitioned. Note that the algorithm M1 is a special case that includes SJF scheduling but which does not involve interval partitioning. To simplify and distinguish it from M1, the symbol M$k$ presented below defaults to $k \geq 2$, thereby ensuring the utilization of the interval partitioning strategy.

Figure 12 portrays differences in task assignments and the number of cores. In algorithms M1 and M$k$, the given tasks have been scheduled based on their respective execution times.

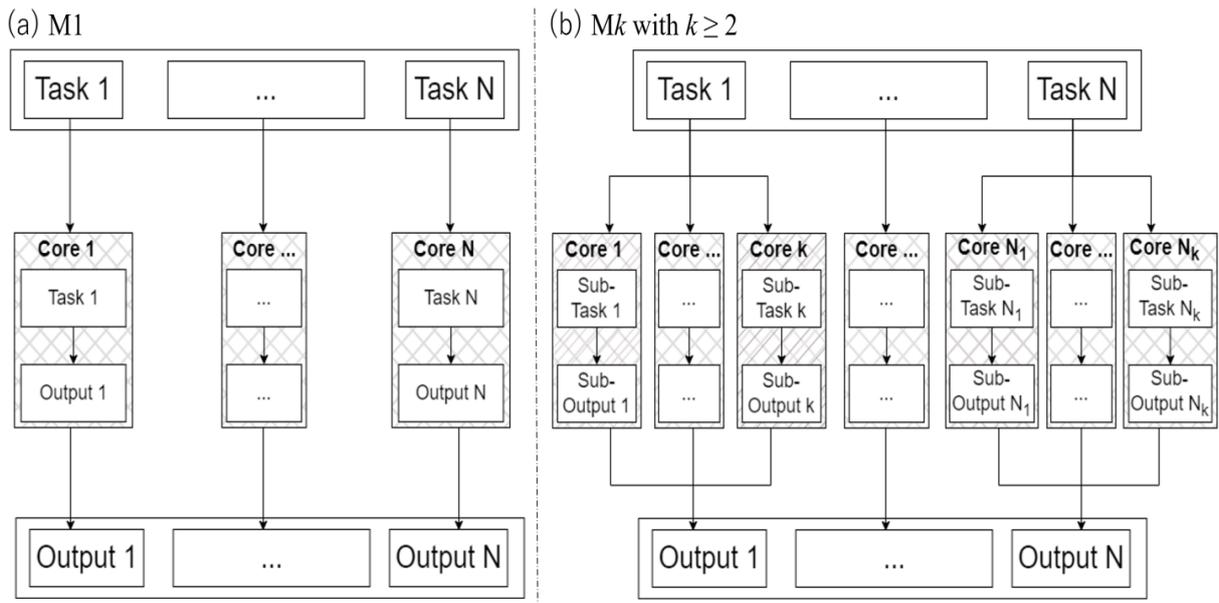

Figure 12. Schematic overview of tasks assigned to multiple cores in the algorithms M1 and M*k*.

To assess the improvement of the proposed parallel algorithms with or without interval partitioning, we have applied the algorithms M1, M2, M4, M8, and M16 on the same dataset. The speedup (*S*) and parallel efficiency (*E*) are computed by varying the number of cores and compared in Figure 13. The comparison indicates that, with an increase in the number of cores, the performances of all the algorithms appear to be improved, but the parallel efficiencies decrease. Comparison between the values of *S* and *E* obtained by executing algorithms M1 and M*k* highlights the effectiveness of the interval partitioning strategy. Based on assessments of computed results, it has been observed that the algorithm M*k*, which incorporates both SJF scheduling and interval partitioning, performs considerably better in terms of speedup and parallel efficiency compared to algorithm M1, with improvements of nearly 20%–30%. Therefore, one can take advantage of such processes to achieve shorter execution times for practical computing applications.

However, it is also found that increasing the number of partitions *k* in algorithm M*k* does not always lead to better performance. This lack of increased performance is attributed to the even distribution of computational load among tasks in the given parallel system, which is formally designated as a load imbalance. In large-scale parallel applications, such a load imbalance results in all processes having to wait for the most overloaded process, here identified by its execution time. This wait introduces a performance penalty because of the idle time, which increases almost linearly as the number of cores increases. Therefore, conducting a preliminary evaluation of the proposed parallel algorithms with regard to load imbalance is particularly notable because their load distribution varies with the numbers of tasks and cores involved.

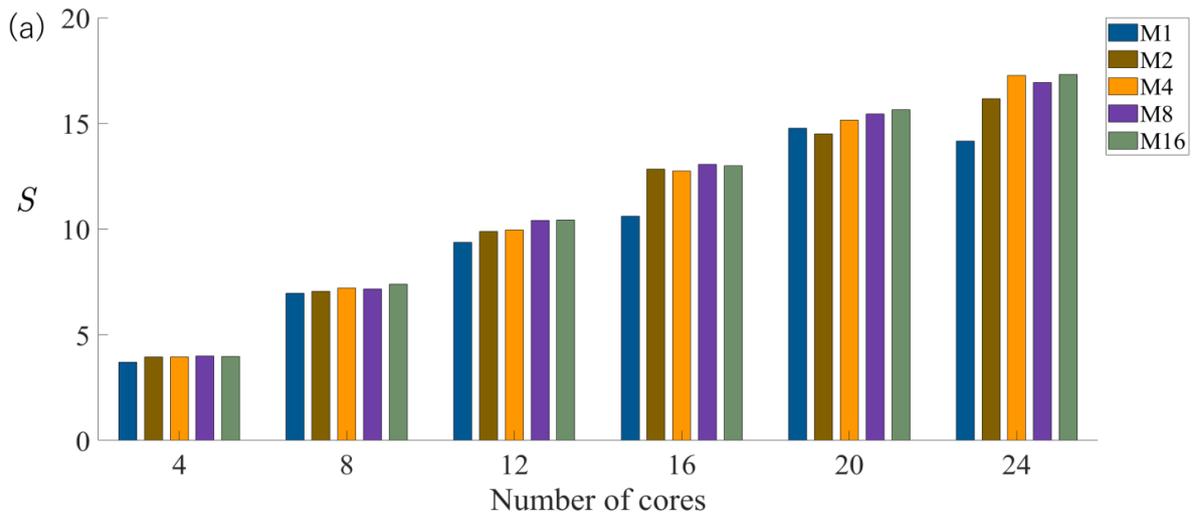

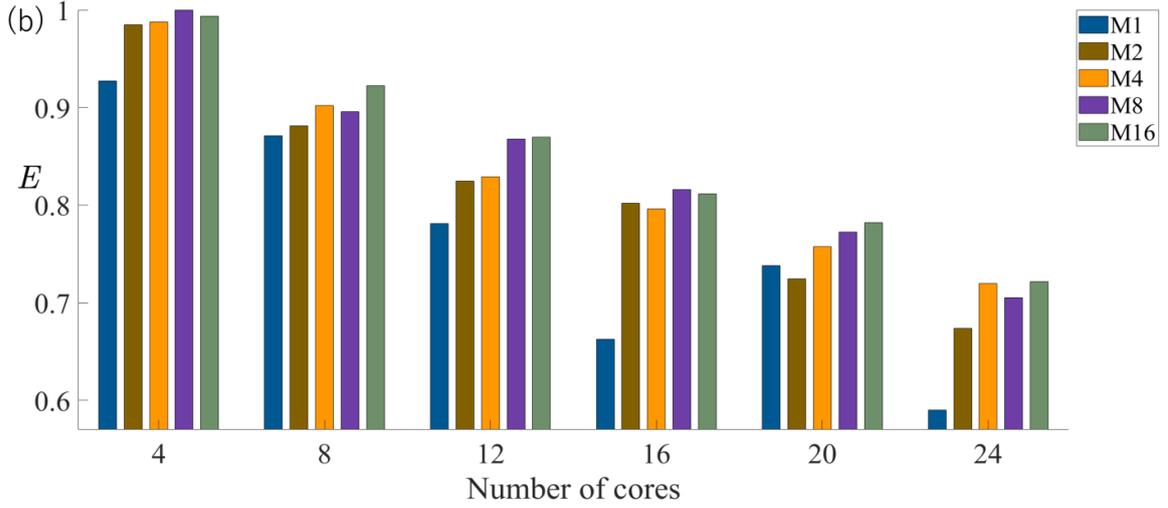

Figure 13. (a) Speedup $S$ and (b) parallel-efficiency $E$ comparisons of parallel implementations using different algorithms and cores. Different colors represent different parallel algorithms.

## 5.3. Statistical metrics

To characterize how unevenly the tasks are distributed, the percent imbalance metric $C_{Loss}$, which is commonly used in this field (Pearce et al. 2012), is employed. It can be represented as

$$C_{Loss} = \left(1 - \frac{\overline{T}}{\max_{n \in [1, p]} T_n}\right) \times 100\%,$$

where $T_n$ stands for the execution time of the $n$-th core, $\overline{T}$ and $\max_{n \in [1, p]} T_n$ respectively represent the mean and maximum execution time over all cores, and $p$ denotes the number of available cores. This metric serves to quantify the performance penalty arising from the load imbalance. However, it lacks the statistical properties of the load distribution, which are necessary for obtaining a comprehensive view of the load distribution, whether the distribution includes a few heavily loaded outliers or numerous slightly imbalanced processes. To address this point, a common statistical moment is used. It is expressed as the following percentage

$$C_{Var} = \sqrt{\frac{1}{p} \sum_{n \in [1, p]} \left(\frac{T_n - \overline{T}}{\overline{T}}\right)^2} \times 100\%,$$

which refers to the degree of variation or inconsistency. Higher $C_{Var}$ values indicate the presence of more infrequent extreme deviations in the current distribution, which might disappear or become less prominent in the subsequent random distribution of parallelized tasks. An ideal distribution would have both $C_{Loss}$ and $C_{Var}$ values of 0, indicating perfect balance and no variation.

To select the algorithm which is best suited for any multicore parallel system, the imbalance metric $C_{Loss}$ and statistical moment $C_{Var}$ are computed and compared for all the subjected algorithms, as depicted in Figure 14. In Figure 14(a), imbalance metric $C_{Loss}$ against the number of cores shows that all algorithms M$k$ (where $k$ = 2, 4, 8, 16) demonstrate comprehensive advantages in all multicore systems and shows that $C_{Loss}$ is maintained consistently below 10% when the number of partitions $k$ in algorithm M$k$ exceeds 4. In Figure 14(b), statistical moment $C_{Var}$ against the number of cores indicates that algorithms M$k$ show marked improvements in all multicore systems, compared with algorithm M1, and that $C_{var}$ is consistently maintained below 3% when the number of partitions $k$ in algorithm M$k$ exceeds 4.

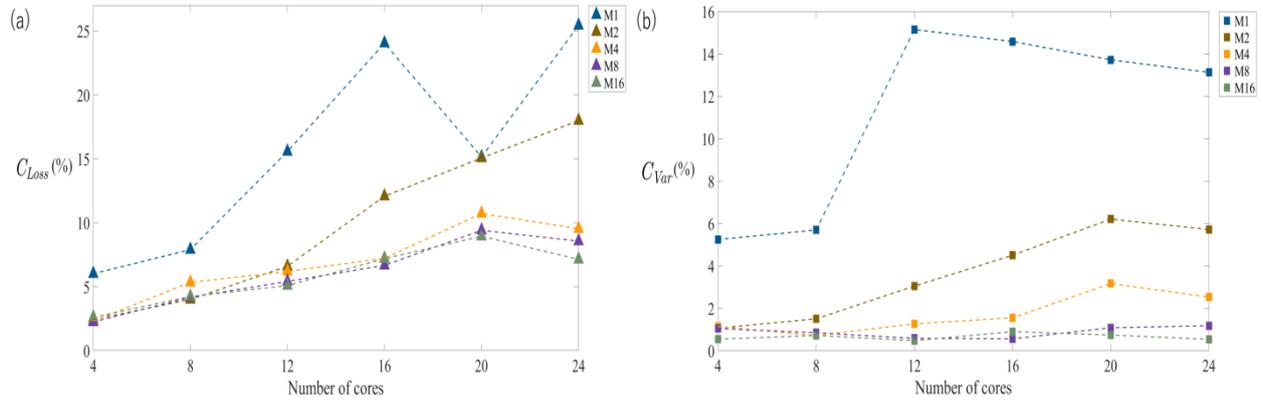

Figure 14. (a) $C_{Loss}$ and (b) $C_{Var}$ against the number of cores computed using different algorithms. Different colors represent different parallel algorithms.

In fact, realistic application environments might encompass diverse and evolving scenarios. Therefore, our parallel algorithm, incorporating SJF scheduling and interval partitioning, is designed to ensure high efficiency, even in scenarios with severe imbalance among the multiple tasks. Specifically, compared to the sequential computation of 30 iterations described at the beginning of this section, which takes over 170 hr, the parallel algorithm M16 accomplishes the same process in about 10 hr on the same personal computer equipped with 24 cores.

## 6. Discussion

The blood circulation network constructed in this study is a closed-loop system, which achieves periodic blood circulation by matching the flow rate and pressure at various bifurcation and confluence junctions and artery–vein connections.

Figure 15 presents the computed variations in blood flow and pressure waves along this network, encompassing arteries, veins, and portal veins. In contrast to the decreases in flow rates resulting from the distribution of blood flow from the aorta to numerous arterial vessels, no great pressure variation among the vascular segments can be recognized. Moreover, compared to arteries, the variations in blood flow and pressure waves within the veins and portal veins are almost indiscernible. These observations are consistent with the physiological characteristics described by Liang et al. (2009).

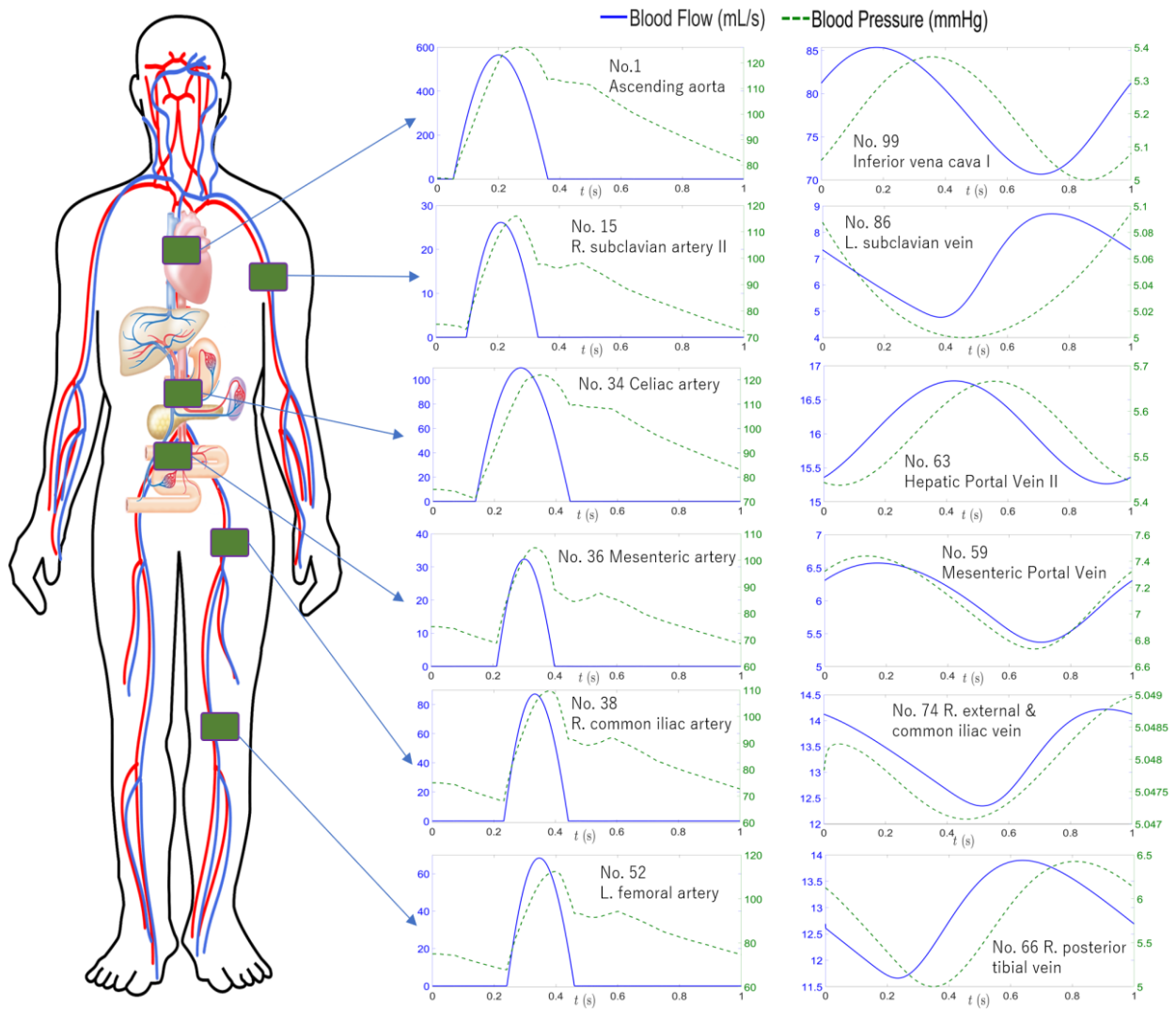

Figure 15. Snapshots of blood flow and pressure wave variations at different locations within the entire blood circulation network, which encompasses the portal venous system. In the left diagram depicting the systemic blood circulation of the whole body, red lines represent arteries, whereas blue lines represent veins and portal

veins. The right subplots show the computed waveforms of blood flow (depicted by blue lines) and blood pressure (shown by green dotted lines) throughout the circulatory system.

The present study, as described in Section 2.2 and as illustrated in Figure 5(b), contributes to an understanding of the flow dynamics in the portal venous system, considering the pathways for substance secretion and absorption reported by Katz et al. (1983) and Koh et al. (2022). Using the computed values of $Q(x, t)$ and $A(x, t)$, an advection–diffusion–reaction equation can be employed to analyze the passive transport of nutrients and hormones. This aspect holds significance from both physiological and medical perspectives.

The parallel algorithms described in Section 5, designed for implementing this computation process, exhibit performance improvement of up to 17 times when compared to sequential computation. However, this marked increase of efficiency in computation is attributed primarily to subsystem $R$, which has a considerably higher time cost than those of other subsystems $F$, $C$, and $H$, as depicted in Table 4. When the tasks are executed, the load imbalance among cores is attributable mainly to the subsystem $R$ execution time. This fact requires us to implement SJF scheduling manually in numerical simulations. Moreover, this situation is not always the case, especially when the blood circulation network involves multiple subsystems with long execution times. For instance, one of the 1D segments can be replaced by a time-consuming three-dimensional model to obtain detailed information about the specific blood vessel. Although the framework provided herein is still applicable to these cases, the performance improvement of parallel computations is expected to be affected by the new execution time corresponding to the alternative model employed in the subsystem. Our future work will include an attempt to implement SJF scheduling automatically by incorporating dynamic load balancing schemes.

In addition, the constructed closed-loop vascular system would be able to serve as a virtual in-silico experimentation platform, allowing independent control over all parameters, including blood vessel geometries and viscoelastic properties. This capability enables the investigation of widely diverse physiological and pathological conditions.

## 7. Conclusion

This study specifically examines the closed-loop modeling of an entire human circulation system using multicore computers. The constructed multiscale model comprises 1D models for the arterial, venous, and portal venous systems, along with 0D lumped-parameter models for the

heart–pulmonary circulation and micro-vasculatures. Regarding the challenge that substantial computational resources are demanded for the global multiscale models of entire blood circulation networks, an iterative procedure is introduced to implement the parallel computations with high efficiency by reducing the execution time within a multicore environment. Moreover, a novel feature of this model is the detailed one-dimensional description of the portal venous system, which holds great potential for future investigation of the mechanisms underlying insulin–glucose dynamics related to the pathological states of diabetes.

## 8. Acknowledgement

This work is supported by JST [Moonshot R&D] [Grant Number JPMJMS2023].

## References

Alastruey J, Parker K, Peiró J, Byrd S, Sherwin S. 2007. Modelling the circle of Willis to assess the effects of anatomical variations and occlusions on cerebral flows. Journal of Biomechanics 40(8):1794–1805.

Anliker M, Rockwell RL, Ogden E. 1971. Nonlinear analysis of flow pulses and shock waves in arteries. Zeitschrift für angewandte Mathematik und Physik ZAMP 22(2):217–246.

Audebert C, Bucur P, Bekheit M, Vibert E, Vignon-Clementel IE, Gerbeau J-F. 2017. Kinetic scheme for arterial and venous blood flow, and application to partial hepatectomy modeling. Computer Methods in Applied Mechanics and Engineering 314:102–125.

Banerjee D, Browne JC. 1996. Complete parallelization of computations: integration of data partitioning and functional parallelism for dynamic data structures. Proceedings of International Conference on Parallel Processing, Honolulu, HI, USA, pp. 354-360.

Barakat M. 2004. Non-pulsatile hepatic and portal vein waveforms in patients with liver cirrhosis: concordant and discordant relationships. The British Journal of Radiology 77(919):547–550.

Barral J-P, Croibier A. 2011. Visceral Vascular Manipulations E-Book. Elsevier Health Sciences.

Berger MJ, Bokhari SH. 1987. A Partitioning Strategy for Nonuniform Problems on Multiprocessors. IEEE Transactions on Computers C-36(5): 570–580.

Cheng CP, Herfkens RJ, Taylor CA. 2003. Inferior vena caval hemodynamics quantified in vivo at rest and during cycling exercise using magnetic resonance imaging. American Journal of Physiology-Heart and Circulatory Physiology 284(4):H1161–H1167.

Chi Z, Beile L, Deyu L, Yubo F. 2022. Application of multiscale coupling models in the numerical study of circulation system. Medicine in Novel Technology and Devices, page 100117.

Formaggia L, Lamponi D, Quarteroni A. 2003. One-dimensional models for blood flow in arteries. Journal of Engineering Mathematics 47(3):251–276.

Formaggia L, Lamponi D, Tuveri M, Veneziani A. 2006. Numerical modeling of 1d arterial networks coupled with a lumped parameters description of the heart. Computer Methods in Biomechanics and Biomedical Engineering, 9(5):273–288.

Formaggia L, Quarteroni A, Veneziani A. 2009. Multiscale models of the vascular system. In Cardiovascular Mathematics, pages 395–446. Springer.

Gorg C, Riera-Knorrenschild J, Dietrich J. 2002. Colour doppler ultrasound flow patterns in the portal venous system. The British Journal of Radiology 75(899):919–929.

He Y, Liu H, Himeno R. 2004. A one-dimensional thermo-fluid model of blood circulation in the human upper limb. International Journal of Heat and Mass Transfer 47(12-13):2735–2745.

Katz LD, Glickman MG, Rapoport S, Ferrannini E, DeFronzo RA. 1983. Splanchnic and peripheral disposal of oral glucose in man. Diabetes 32(7):675–679.

Koh H-CE, Cao C, Mittendorfer B. 2022. Insulin clearance in obesity and type 2 diabetes. International Journal of Molecular Sciences 23(2).

Levick JR. 2013. An introduction to cardiovascular physiology. Butterworth–Heinemann.

Liang F, Takagi S, Himeno R, Liu H. 2009. Biomechanical characterization of ventricular–arterial coupling during aging: a multi-scale model study. Journal of Biomechanics 42(6):692–704.

Milišić V, Quarteroni A. 2004. Analysis of lumped parameter models for blood flow simulations and their relation with 1d models. ESAIM: Mathematical Modelling and Numerical Analysis 38(4):613–632.

Müller LO, Toro EF. 2014. A global multiscale mathematical model for the human circulation with emphasis on the venous system. International Journal for Numerical Methods in Biomedical Engineering 30(7):681–725.

Müller LO, Watanabe SM, Toro EF, Feijóo RA, Blanco PJ. 2023. An anatomically detailed arterial—venous network model. Cerebral and coronary circulation. Frontiers in Physiology 14:1162391.

Murgo JP, Westerhof N, Giolma JP, Altobelli SA. 1980. Aortic input impedance in normal man: relationship to pressure wave forms. Circulation 62(1):105–116.

Mynard JP. Smolich JJ. 2015. One-dimensional haemodynamic modeling and wave dynamics in the entire adult circulation. Annals of Biomedical Engineering 43(6):1443–1460.

Pearce O, Gamblin T, De Supinski BR, Schulz M, Amato NM. 2012. Quantifying the effectiveness of load balance algorithms. In Proceedings of the 26th ACM International Conference on Supercomputing, 185–194.

Quarteroni A, Formaggia L. 2004. Mathematical modelling and numerical simulation of the cardiovascular system. Handbook of Numerical Analysis 12:3–127.

Quarteroni A, Lassila T, Rossi S, Ruiz-Baier R. 2017. Integrated heart—coupling multiscale and multiphysics models for the simulation of the cardiac function. Computer Methods in Applied Mechanics and Engineering 314:345–407.

Quarteroni A, Manzoni A, Vergara C, et al. 2019. Mathematical modelling of the human cardiovascular system: data, numerical approximation, clinical applications, volume 33. Cambridge University Press.

Quarteroni A, Valli A. 2008. Numerical approximation of partial differential equations, volume 23. Springer Science & Business Media.

Regazzoni F, Salvador M, Africa PC, Fedele M, Dede L, Quarteroni A. 2022. A cardiac electromechanical model coupled with a lumped-parameter model for closed-loop blood circulation. Journal of Computational Physics 457:111083.

Sherwin SJ, Formaggia L, Peiro J, Franke V. 2003. Computational modelling of 1d blood flow with variable mechanical properties and its application to the simulation of wave propagation in the human arterial system. International Journal for Numerical Methods in Fluids 43(6-7):673–700.

Sun Y, Beshara M, Lucariello RJ, Chiaramida SA. 1997. A comprehensive model for right-left heart interaction under the influence of pericardium and baroreflex. American Journal of Physiology-Heart and Circulatory Physiology 272(3):H1499–H1515.

Toro EF, Celant M, Zhang Q, Contarino C, Agarwal N, Linninger A, Müller LO. 2022. Cerebrospinal fluid dynamics coupled to the global circulation in holistic setting: mathematical models, numerical methods and applications. International Journal for Numerical Methods in Biomedical Engineering 38(1):e3532.

Vignesh V, Kumar KS, Jaisankar N. 2013. Resource management and scheduling in cloud environment. International Journal of Scientific and Research Publications 3(6):1-6.

Wolf RL, King B, Torres VE, Wilson DM, Ehman R. 1993. Measurement of normal renal artery blood flow: cine phase-contrast MR imaging vs. clearance of p-Aminohippurate. AJR. American Journal of Roentgenology 161(5):995–1002.

Zitnik RS, Rodich FS, Marshall HW, Wood EH. 1965. Continuously recorded changes of thoracic aortic blood flow in man in response to leg exercise in supine position. Circulation Research 17(2):97–105.